\documentclass[aps,prl,reprint,preprintnumbers,superscriptaddress,amsmath,amssymb,bibnotes,longbibliography]{revtex4-2}
\usepackage{dcolumn}
\usepackage{bm}
\usepackage{graphicx}
\usepackage{amssymb}
\usepackage{amsmath}
\usepackage{epstopdf}
\usepackage{booktabs}
\usepackage{makecell} 
\usepackage[pdfborder=001,colorlinks,linkcolor=blue,anchorcolor=blue,citecolor=blue,urlcolor=blue,filecolor=blue,menucolor=blue,runcolor=blue]{hyperref}

\begin{document}
\title{G-type antiferromagnetic structure in Rb$_{1-\delta}$V$_2$Te$_2$O}
\author{Wu Xie}
\affiliation{Spallation Neutron Source Science Center, Dongguan 523803, P. R. China}
\affiliation{Institute of High Energy Physics, Chinese Academy of Sciences, Beijing 100049, P. R. China}
\author{Changchao Liu}
\affiliation{School of Physics, Zhejiang University, Hangzhou, China}
\author{Fayuan Zhang}
\affiliation{Quantum Science Center of Guangdong-Hong Kong-Macao Greater Bay Area, Shenzhen 518045, China}
\author{Zhenhong Tan}
\affiliation{Spallation Neutron Source Science Center, Dongguan 523803, P. R. China}
\affiliation{Institute of High Energy Physics, Chinese Academy of Sciences, Beijing 100049, P. R. China}
\author{Wenhai Ji}
\affiliation{Spallation Neutron Source Science Center, Dongguan 523803, P. R. China}
\affiliation{Institute of High Energy Physics, Chinese Academy of Sciences, Beijing 100049, P. R. China}
\author{Nan Zhao}
\affiliation{Spallation Neutron Source Science Center, Dongguan 523803, P. R. China}
\affiliation{Institute of High Energy Physics, Chinese Academy of Sciences, Beijing 100049, P. R. China}
\author{Lingxiang Bao}
\affiliation{Spallation Neutron Source Science Center, Dongguan 523803, P. R. China}
\affiliation{Institute of High Energy Physics, Chinese Academy of Sciences, Beijing 100049, P. R. China}
\author{Dong Zhang}
\affiliation{Spallation Neutron Source Science Center, Dongguan 523803, P. R. China}
\affiliation{Institute of High Energy Physics, Chinese Academy of Sciences, Beijing 100049, P. R. China}
\author{Feiran Shen}
\affiliation{Spallation Neutron Source Science Center, Dongguan 523803, P. R. China}
\affiliation{Institute of High Energy Physics, Chinese Academy of Sciences, Beijing 100049, P. R. China}
\author{Lunhua He}
\affiliation{Spallation Neutron Source Science Center, Dongguan 523803, P. R. China}
\affiliation{Beijing National Laboratory for Condensed Matter Physics, Institute of Physics, Chinese Academy of Sciences, Beijing 100190, China}
\author{Hao Wang}
\affiliation{Spallation Neutron Source Science Center, Dongguan 523803, P. R. China}
\affiliation{Institute of High Energy Physics, Chinese Academy of Sciences, Beijing 100049, P. R. China}
\author{Rong Du}
\affiliation{Spallation Neutron Source Science Center, Dongguan 523803, P. R. China}
\affiliation{Institute of High Energy Physics, Chinese Academy of Sciences, Beijing 100049, P. R. China}
\author{Guanghan Cao}
\email{ghcao@zju.edu.cn}
\affiliation{School of Physics, Zhejiang University, Hangzhou, China}
\author{Chaoyu Chen}
\email{chenchaoyu@sslab.org.cn}
\affiliation{Songshan Lake Materials Laboratory, Dongguan, China.}
\author{Ping Miao}
\email{miaoping@ihep.ac.cn}
\affiliation{Spallation Neutron Source Science Center, Dongguan 523803, P. R. China}
\affiliation{Institute of High Energy Physics, Chinese Academy of Sciences, Beijing 100049, P. R. China}

\date{\today}

\begin{abstract}
Altermagnetism, known for its non-relativistic spin-split band structures with yet compensated moments, is being intensively investigated. Discovering new altermagnetic materials with  characteristics suitable for practical use remains an important ongoing task. Recently a metallic room-temperature altermagnet candidate Rb$_{1-\delta}$V$_2$Te$_2$O with a layered structure and $d$-wave spin symmetry has been reported based on experimental results from the spin-resolved photoemission spectroscopy and scanning tunnelling microscopy/spectroscopy measurements. Here we report neutron powder diffraction investigations on the magnetic structure of Rb$_{1-\delta}$V$_2$Te$_2$O, which show a G-type antiferromagnetic structure below the transition temperature of 337 K. The result is different from the original theoretical expectation, which might lead to new insights on the physics of this altermagnet candidate.

\begin{description}
\item[PACS number(s)]

\end{description}
\end{abstract}

\maketitle

\section{I. Introduction}
Altermagnetism represents one of typical unconventional mechanism for spin-polarized electronic states realization, in addition to the Rashba–Dresselhaus effect arising from the spin–orbit coupling (SOC) in compounds with inversion symmetry breaking, and Zeeman-type spin splitting in ferromagnets with time-reversal symmetry breaking. While extensive theoretical and experimental efforts have led to the discovery of a series of altermagnets with typical characteristics \cite{RuO2-1,RuO2-2,RuO2-3,RuO2-4,v221,AM_PRX, MnTe-1,MnTe-2,KVTO,RbVTO,CPL}, this field is confronting controversies on specific candidates \cite{review1,review2,arpes_AM}. Exploring new altermagnetic materials with favorable properties for practical use remains an important ongoing task.

Recently a metallic room-temperature altermagnet candidate Rb$_{1-\delta}$V$_2$Te$_2$O has been reported, which features a layered structure and exhibits spin polarization of $d$-wave symmetry \cite{RbVTO}. Such $d$-wave symmetric spin splitting is favorable for generating non-relativistic spin currents, meanwhile the layered structure facilitates the exfoliation from bulk into thin films and the integration with other materials in spintronic device fabrications. With its high transition temperature above room temperature, Rb$_{1-\delta}$V$_2$Te$_2$O serves as one of the most promising altermagntic material platform for both fundamental research and applications in spintronics and valleytronics devices. 

\begin{figure}[htb]
     \begin{center}
     \includegraphics[width=1.0\columnwidth,keepaspectratio]{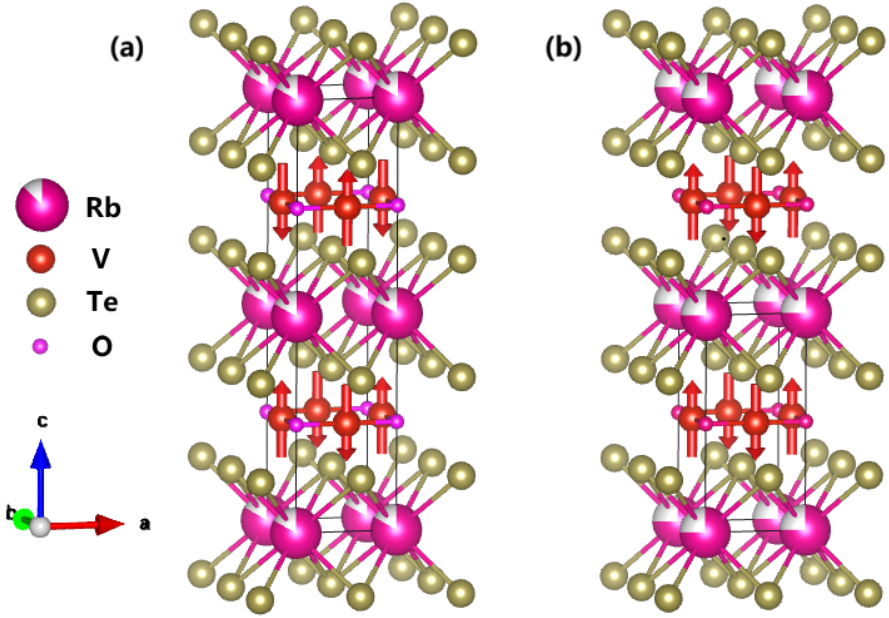}
     \end{center}
     \caption{(Color online) The illustration of two typical magnetic structures of Rb$_{1-\delta}$V$_2$Te$_2$O. The G-type and C-type AFM structures are displayed in panel (a) and (b), respectively.  }
     \label{fig1}
\end{figure}

Experimentally, although the angle-resolved photoemission spectroscopy (ARPES) and scanning tunnelling microscopy/spectroscopy (STM/STS) measurements \cite{RbVTO} show opposite spin splitting between crystal-symmetry-paired valleys and suppression of intervalley scattering as a consequence of the spin–valley locking, respectively, investigations on the microscopic magnetic structure in Rb$_{1-\delta}$V$_2$Te$_2$O have not been reported. The calculations based on density functional theory (DFT) have predicted a ground state of C-type antiferromagnetic (AFM) structure, whose energy is only around 1 meV/supercell lower than the G-type AFM structure in Rb$_{1-\delta}$V$_2$Te$_2$O, as shown in Fig.~\ref{fig1} \cite{SI_RbVTO} . Very recently, in the sister compound Cs$_{1-\delta}$V$_2$Te$_2$O, a G-type AFM structure has been confirmed based on neutron diffraction investigations, where similar spin polarization has been observed on the spin-resolved ARPES measurements \cite{CsVTO}. The concepts of ``hidden altermagnetism'' and local altermagnetic order were adopted to reconcile the obtained magnetic structure and the spin slittings in momentum space \cite{CsVTO}, as proposed earlier by theorists \cite{HM}. Since G-type AFM structure favors altermagnetism only in a single sector in-plane while C-type AFM structure is favorable for altermagnetism as a whole, it is rather important to pin down the microscopic magnetic structure in Rb$_{1-\delta}$V$_2$Te$_2$O, which poses strong restrictions on the interpretation of the observed $d$-wave symmetric spin polarization.

In this work, we report the neutron powder diffraction (NPD) investigations on the magnetic structure of Rb$_{1-\delta}$V$_2$Te$_2$O ($\delta \sim$ 0.88). Pure magnetic peaks indexed by a magnetic propagation vector of (0, 0, 0.5) were observed below the N$\rm\acute{e}$el temperature $T\rm_N$ of 337 K, ruling out the possibility of C-type AFM structure. Rietveld refinements using FullProf software suite unambiguously point to a G-type AFM structure in Rb$_{1-\delta}$V$_2$Te$_2$O.

\section{II. Experimental methods}
Polycrystals of Rb$_{1-\delta}$V$_2$Te$_2$O were synthesized via a solid-state reaction method, as reported in \cite{RVTO_stru}. The as-prepared sample was characterized by powder x-ray diffractions (XRD) at room temperature using a PANalytical diffractometer (Empyrean Series 2) with a monochromatic Cu-K$_\alpha1$ radiation. A minor impurity phase of V$_2$O$_3$ less than 5~$\%$ was determined from XRD. Magnetic susceptibility was measured on a Quantum Design Physical Property Measurement System (QD-PPMS) using a Vibrating Sample Magnetometer (VSM) insertion.

Neutron powder diffraction (NPD) measurements was conducted at the time-of-flight high-resolution neutron diffractometer (TREND) and
the general-purpose powder diffractometer (GPPD) at the China Spallation Neutron Source (CSNS) \cite{trend1,trend2,trend3,trend4}. The well-grinded powder sample was loaded into a 10 mm VNi holder under Helium atmosphere and sealed via Indium wire. The sample was mounted in a cryofurnace and the NPD data were collected at temperatures ranging from 5 K to 350 K, across the N$\rm\acute{e}$el temperature. Due to the air- and water- sensitiveness of the sample, all experimental operations related to sample were conducted under inert gases. 

The Rietveld refinements of Neutron data were performed using FullProf software suite \cite{Fullprof}, where the $K\rm_{SEARCH}$ program was used for determining the magnetic propagation vector ($k$-vector), and BasIreps was adopted for the irreducible representation (Irreps) analysis during the magnetic model construction. In addition, programs on Bilbao Crystallographic Server (https://www.cryst.ehu.es), in particular MAXMAGN and k-SUBGROUPSMAG, were adopted to identify the maximal magnetic space groups \cite{Bilbao}. Finally, VESTA was utilized for structure visualization and presentation \cite{VESTA}.

\begin{figure}[htb]
\centering\includegraphics[width=0.95\columnwidth]{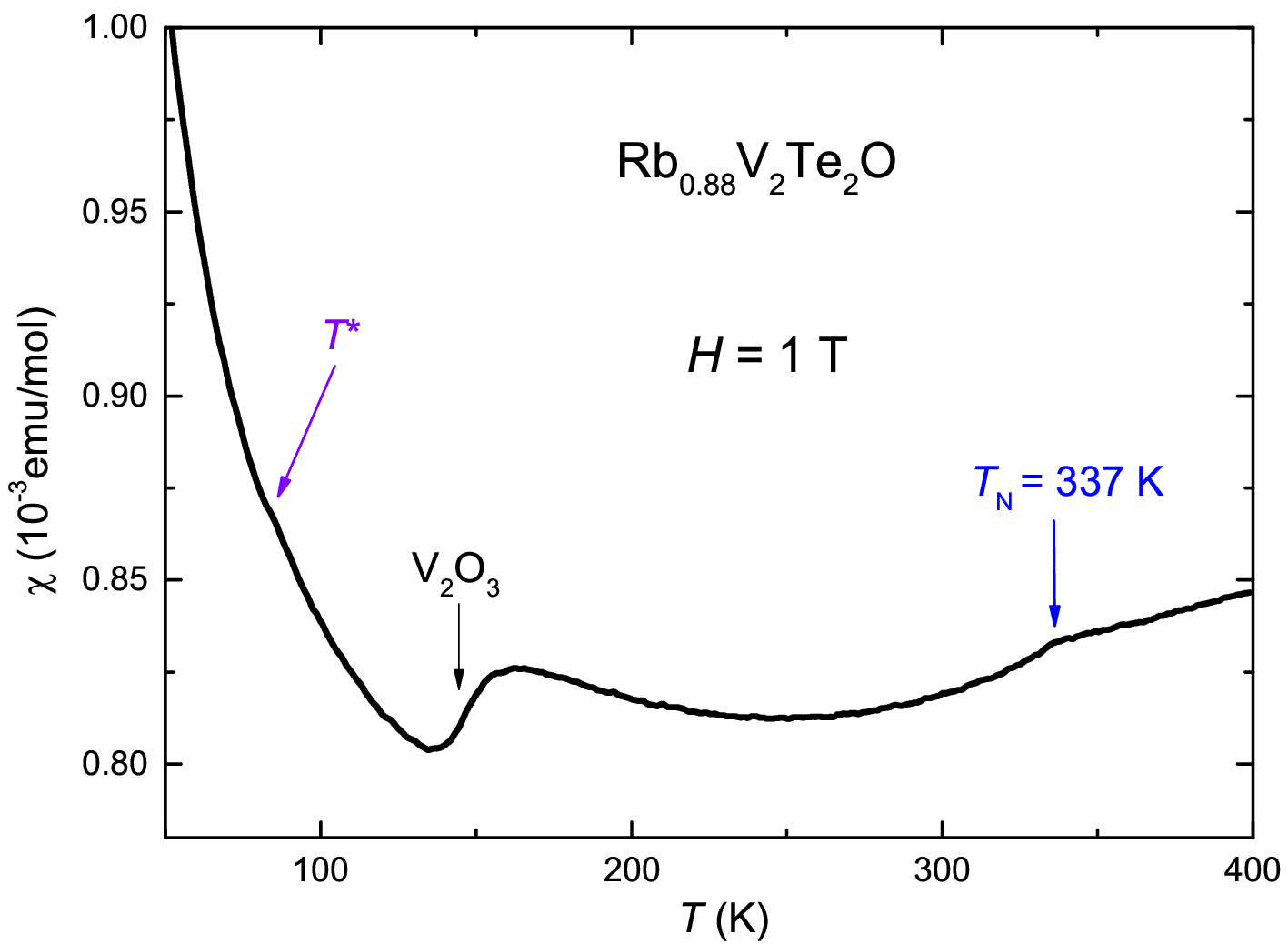}
\caption{(Color online) Magnetic susceptibility $\chi$ versus $T$ of  Rb$_{1-\delta}$V$_2$Te$_2$O polycrystal under a magnetic field of 1 T, where the positions of two transitions $T_\mathrm{N}$ and $T^*$ are noted by the blue and purple arrows, respectively. The drop at around 150 K are related to the structural transition of V$_2$O$_3$. }
\label{fig4}
\end{figure}

 \begin{figure}[b]
\centering\includegraphics[width=1.0\columnwidth,keepaspectratio]{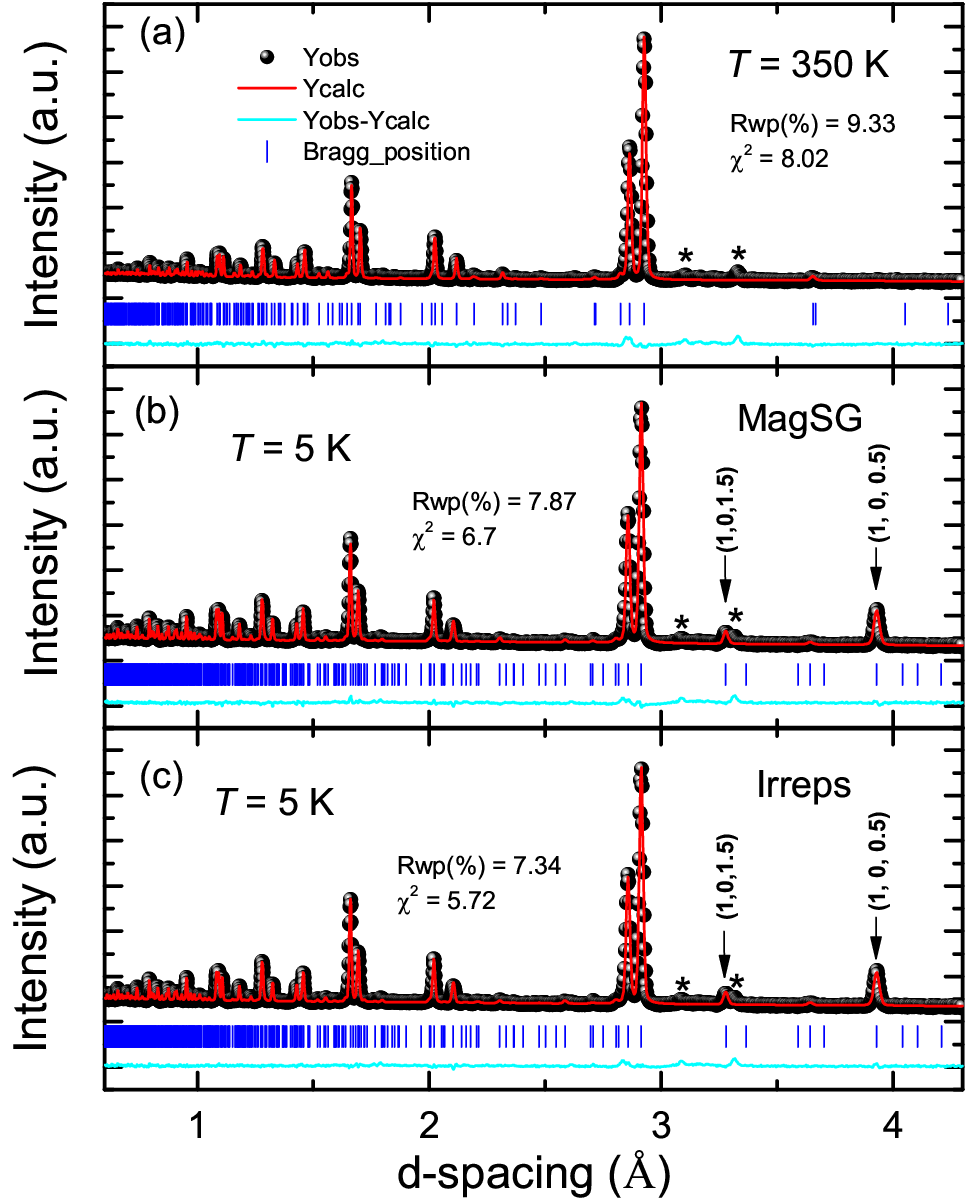}
\caption{(Color online) The results of Rietveld refinements of the NPD data obtained from TREND at (a) $T = 350 $ K, and (b)(c) $T = 5$ K, respectively. Panel (b) and (c) show the 5 K data fitted with the magnetic phase modeled using two different methodologies, respectively, which result in the same G-type AFM structure.  The asterisks denote the peaks from a minor impurity phase, while the black arrows in panels (b) and (c) indicate the position of two strong magnetic peaks. }
\label{fig2}
\end{figure}

\begin{table}[htbp]
\centering
\caption{The G-type Antiferromagnetic  structure of Rb$_{1-\delta}$V$_2$Te$_2$O is described under its magnetic space group 
(MagSG, crystallographic description), with basic information about its relation to 
its parent paramagnetic structure. All items in the table are supported by the 
magnetic CIF format (mcif).}
\label{tab1}

\begin{tabular}{l | l} 
\toprule 
Items & Details  \\
\midrule 
Compound &   Rb$_{1-\delta}$V$_2$Te$_2$O\\

Parent space group &  $P4/mmm$ (No.123); \\

Propagation vector(s) & (0, 0, 0.5); \\
\midrule 
\makecell[l]{Basis transformation \\ from parent basis} &  (a,b,2c; 0,0,0 );\\
\midrule 
MagSG symbol (number) &   $P_c$$4_2/mcm$ (132.456) \\
\midrule 
\makecell[l]{Basis transformation \\ to standard setting of MagSG} &  (a,b,c; 0,0,1/4);\\
\midrule 
Unit cell parameters (\text{\AA}) &   \makecell[l]{a = 4.03813(8) \\
b = 4.03724(16) \\
c = 16.8181(4)\\
$\alpha$ = 90.0000\\
$\beta$ = 90.0000\\
$\gamma$ = 90.0000}\\
\midrule 
MagSG symmetry operations &   \makecell[l]{ 1 x,y,z,+1 \\
   2  -x,-y,-z,-1 \\
   3  y,-x,z,-1      \\
   4  y,x,-z,-1        \\
   5  -y,-x,z,+1    \\
   6  -x,y,z,-1      \\
   7  y,x,z,+1     \\
   8  x,-y,z,-1\\
   9  x,y,-z,-1\\
  10  -y,x,z,-1\\
  11  -y,-x,-z,-1\\
  12  y,-x,-z,+1\\
  13  x,-y,-z,+1\\
  14  -x,y,-z,+1\\
  15  -x,-y,z,+1\\
  16  -y,x,-z,+1}         \\
\midrule 
\makecell[l]{MagSG symmetry centering \\operations} & \makecell[l]{ 1  x,y,z,+1\\
   2  x,y,z+1/2,-1}    \\
\midrule 
Positions of atoms & \makecell[l]{ Te1 0.50000  0.50000  0.13539 \\
 V1  0.00000  0.50000  0.25000 \\
 O1 0.00000  0.00000  0.25000  \\
 Rb1 0.00000  0.00000  0.00000  }                    \\
\midrule 
\makecell[l]{Magnetic moment ($\mu\rm_B$) \\ of magnetic 
atoms, \\  symmetry constraints\\  and moment 
magnitudes \\ at 5 K} &  \makecell[l]{ V1  0.00000  0.00000  1.425(13) \\ 
(0, 0, $m_z$) \\ 
1.425(13)}       \\
\bottomrule 
\end{tabular}

\end{table}

\section{III. Results and discussion}

The polycrystalline sample for neutron study has been firstly checked with magnetic susceptibility $\chi(T)$ measurement. As shown in Fig.~\ref{fig4}, two transitions have been observed: the magnetic transition $T_\mathrm{N}$ at 337 K and the metal-to-metal transition $T^*$ at 116 K. The two transitions are both higher than that reported in literature \cite{RbVTO,RVTO_stru}, which are most likely due to the sample dependence, such as the slightly different Rb-occupancy.

Figure~\ref{fig2} shows the FullProf refinement results of powder neutron diffraction data of Rb$_{1-\delta}$V$_2$Te$_2$O above and below $T\rm_N$.
As shown in Fig.~\ref{fig2}(a), for $T = 350$ K, a crystal structure model of Rb$_{1-\delta}$V$_2$Te$_2$O with space group (SG) $P4/mmm$ (No.123) is adopted for the refinement, together with the high temperature phase of V$_2$O$_3$ (SG: $R$-3$c$) as a known impurity phase. Besides, there are at least one more impurity peak as indicated by asterisks in Fig.~\ref{fig2}(a). The peaks from impurity phases show negligible temperature dependence down to 5 K. The refined crystal structure for Rb$_{1-\delta}$V$_2$Te$_2$O is consistent with that reported in literature \cite{RVTO_stru}. The obtained Rb occupancy is about 0.88, corresponding to a $\delta$ of 0.12 in Rb$_{1-\delta}$V$_2$Te$_2$O. 

Upon cooling, two additional peaks are observed as denoted by the arrows in Fig.~\ref{fig2}(b) and (c). Using $K\rm_{SEARCH}$, a $k$-vector of (0, 0, 0.5) is undoubtedly determined, and the two magnetic peaks can be indexed as (1, 0, 0.5) and (1, 0, 1.5), respectively.  The G-type AFM and C-type AFM structures as shown in Fig.~\ref{fig1} are indexed by different magnetic propagation vectors: (0, 0, 0.5) for G-type, and (0, 0, 1) for C-type. Consequently, the G-type AFM structure will generate additional reflections (magnetic peaks) below the N$\rm\acute{e}$el temperature $T\rm_N$, while for C-type AFM structure, the magnetic peaks all appear on top of the nuclear peaks. Therefore, the observation of additional magnetic peaks rules out the possibility of the C-type AFM structure.

Below $T_\mathrm{N}$, a magnetic model is further considered in the FullProf refinement through two methodologies: magnetic space group (MagSG) and irreducible representation (Irreps). The maximal magnetic space groups compatible with a $k$-vector of (0, 0, 0.5) are analyzed using MAXMAGN on Bilbao Crystallographic Server. While in total 12 maximal magnetic space groups are listed, only four among them allows nonzero magnetic moments for the V-ions: $P_c4_2/mcm$ ($\#$132.456), $P_c4/mcc$ ($\#$124.360), $C_cmcm$ ($\#$63.466) and $P_amma$ ($\#$51.298). Among them, $P_c4_2/mcm$ and $P_c4/mcc$ correspond to G-type and A-type AFM structure, respectively, with moments along the $c$-axis, while the moments are constrained in the $ab$-plane for $C_cmcm$ and $P_amma$. We have tested all the four magnetic models in FullProf refinements, and finally concluded that the G-type AFM model of magnetic space group $P_c4_2/mcm$ ($\#$132.456) gives the best fit to the magnetic peaks, as shown in Fig.~\ref{fig2}(b). As for the representation analysis, three Irreps are obtained, $i.e.$, $\Gamma_2$ of one dimension, $\Gamma_4$ of one dimension, and 2$\Gamma_9$ of two dimensions (1$\Gamma_2$ + 1$\Gamma_4$ + 2$\Gamma_9$), which correspond to G-type AFM, A-type AFM and in-plane noncollinear AFM structures, respectively. The Rietveld refinement also indicate that the G-type AFM model of $\Gamma_2$ fits best to the pattern as shown in Fig.~\ref{fig2}(c), consistent with the results from the MagSG methodology. The obtained V-moment size at 5 K is 1.425(13)$\mu_B$ from MagSG and 1.449(12)$\mu_B$ from Irreps, respectively. The basic information of the obtained G-type AFM structure of Rb$_{1-\delta}$V$_2$Te$_2$O, described under its MagSG, are listed in Table~\ref{tab1}.
Note that the impurity phase V$_2$O$_3$ undergoes a first-order structural transition at $\sim$170 K \cite{V2O3}. Therefore, for the refinements below 170 K, the low temperature phase of V$_2$O$_3$ (SG: $C2/c$) is included in the Rietveld refinement. Note also that three batches of polycrystalline sample have been measured at two different neutron diffractometers (TREND and GPPD) at CSNS. All results conclude the same G-type AFM structure in Rb$_{1-\delta}$V$_2$Te$_2$O. 

\begin{figure}[htb]
\begin{center}
     \includegraphics[width=\columnwidth,keepaspectratio]{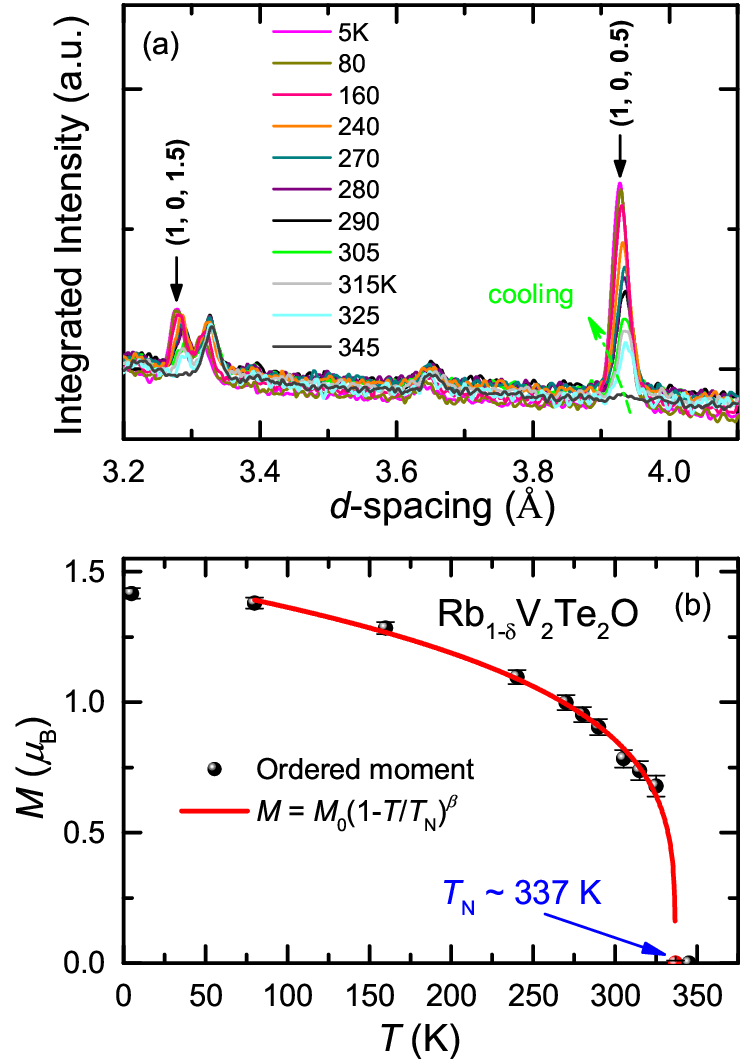}
     \end{center}
     \caption{(Color online) (a) NPD data obtained from TREND in a selected $d$-range measured at a series of temperatures, showing the temperature dependent magnetic peak intensity. (b) Temperature dependence of the ordered moment. The red solid line is the fitted result as described in the main text. The red  point is the transition point obtained from the temperature dependent magnetic susceptibility measurement. }
\label{fig3}
\end{figure}

Fig.~\ref{fig3}(a) shows the temperature dependence of the two magnetic peaks, whose intensities increase as temperature decreases. The temperature dependence of the magnetic moment size obtained from Rietveld refinement are plotted in Fig.~\ref{fig3}(b). An empirical formula $M = M_0 (1-T/T_\mathrm{N})^\beta$ is adopted to fit the temperature dependence of the ordered moments as shown in Fig.~\ref{fig3}(b), where $M_0$ is the ordered moment at zero-temperature and $\beta$ is the critical exponent.  The obtained $\beta$ value is $\sim$0.25 $\pm$ 0.00746, which is much smaller than the mean-field theory prediction ($\beta$ = 0.5). In the sister compound KV$_2$Te$_2$O \cite{KVTO_neutron}, $\beta$ was fitted to be about 0.48$\pm$0.02, where the magnetic moments were derived from the temperature dependent magnetic peak intensity via scaling.

\section{IV. Discussion and Summary}

In summary, we have performed NPD investigations on the $d$-wave altermagnet candidate Rb$_{1-\delta}$V$_2$Te$_2$O and determined a G-type AFM structure with moments along the $c$-axis (Fig.~\ref{fig1}(a) and Table~\ref{tab1}) for the ground state. The result is different from the original expectation, where a C-typed AFM structure is expected to support a global $d$-wave altermagnetism. Since ARPES and STM data have revealed concrete experimental evidences for the $d$-wave spin symmetry, analogous to the recent report for Cs$_{1-\delta}$V$_2$Te$_2$O \cite{CsVTO}, a hidden altermagnetism with alternating local spin polarization arising from single stacking layer of the G-type AFM structure might act as a proper scenario to reconcile all experimental discoveries.

\section{V. Acknowledgments}
\begin{acknowledgments}
This work is supported by National Key R$\&$D Program of China under Contracts (Grant No. 2023YFE0105700, and No. 2023YFA1406101), the National Natural Science Foundation of China (NSFC) (Grant No. 12304183, No. 12574068, and No. 12504185), Guangdong Innovative $\&$ Entrepreneurial Research Team Program (No.2021ZT09C539), and the Guangdong Basic and Applied Basic Research Foundation (Grant No. 2022B1515120014, No. 2023B0303000003 and No. 2023B1515120060). The authors appreciate the neutron beamtime at the High-resolution Neutron Diffractometer (TREND) (https://cstr.cn/31113.02.CSNS.TREND) and General Purpose Powder Diffractometer (GPPD) (https://csns.cn/31113.02.CSNS.GPPD) at the China Spallation Neutron Source (CSNS) (https://cstr.cn/31113.02.CSNS). The authors also thank Dr. Wanju Luo for his assistance in the experiment.
\end{acknowledgments}

\end{document}